# A Crowdsourcing Procedure for the Discovery of Non-Obvious Attributes of Social Image

Mark Melenhorst[1], María Menendez Blanco[2], Martha Larson[1]
[1]Multimedia Computing, Department of Intelligent Systems, Delft University of Technology
[2] Department of Information Engineering and Computer Science, University of Trento
m.s.melenhorst@tudelft.nl, menendez@disi.unitn.it, m.a.larson@tudelft.nl

## ABSTRACT

Research on mid-level image representations has conventionally concentrated relatively obvious attributes and overlooked *non-obvious attributes*, i.e., characteristics that are not readily observable when images are viewed independently of their context or function. Non-obvious attributes are not nessarily easily nameable, but nonetheless they play a systematic role in people's interpretation of images. Clusters of related non-obvious attributes, called *interpretation dimensions*, emerge when people are asked to compare images, and provide important insight on aspects of social images that are considered relevant. In contrast to aesthetic or affective approaches to image analysis, non-obvious attributes are not related to the personal perspective of the viewer. Instead, they encode a conventional understanding of the world, which is tacit, rather than explicitly expressed. This paper introduces a procedure for discovering non-obvious attributes using crowdsourcing. We discuss this procedure using a concrete example of a crowdsourcing task on Amazon Mechanical Turk carried out in the domain of fashion. An analysis comparing discovered non-obvious attributes with user tags demonstrated the added value delivered by our procedure.

## 1. INTRODUCTION

This paper contributes to the forward progress of automatic content-based image indexing by pointing out an overlooked aspect of how people interpret social images, and presenting a method that will allow it to be understood and addressed by the multimedia content analysis community. Specifically, we introduce a procedure for eliciting *non-obvious attributes* to describe social images. We define non-obvious attributes as: *characteristics that people find important when they are focused on context or function.* Note that non-obvious attributes can describe the entities and events depicted in images, but they can also describe the images themselves. "Context" refers to *when* something is used, i.e., a set of conventional circumstances, and "function" refers to *how* something is used, i.e., its purpose. We refer to the characteristics as *non-obvious* because they do not readily jump to mind when the images are examined outside of an appropriate context, or independently of any considerations of function.

The notion of non-obvious attributes is illustrated by the three sets of images in Figure 1. The dogs in the top row can be distinguished by obvious characteristics such as size and color. Non-obvious characteristics go beyond these attributes. An example of a non-obvious attribute that distinguishes these dogs is "handle with care". It is possible to describe the dogs in terms of whether the phrase "handle with care" applies.

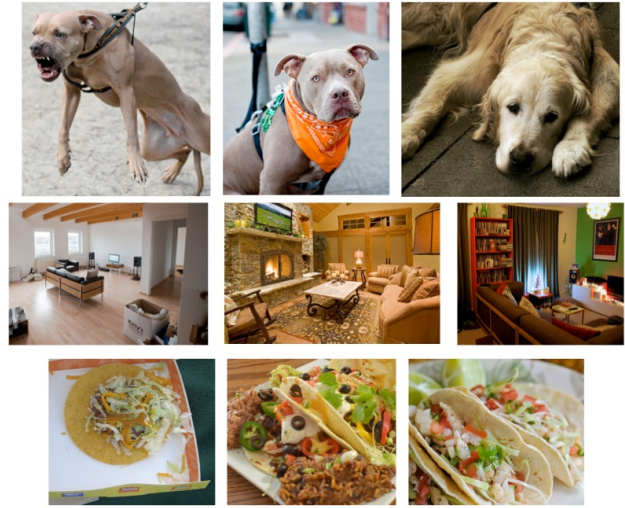

**Figure 1:** Examples of non-obvious attributes belonging to interpretation dimensions for social images. Each row of images depicts the same visual concepts. Top: an example interpretation dimension for 'dog' runs from 'handle with care' to 'companionable'; Middle: an example dimension for 'livingroom' runs from 'minimalist' to 'homey/homely'. Bottom: an example dimenseion for 'tacos' runs from 'fast food' to 'home style'.

This example reveals two properties of non-obvious attributes that make them challenging to study. These properties help to explain why they have been, up to this point, overlooked in the research community. First, non-obvious attributes are difficult to name. The idea of "handle with care" associated with a dog cannot be described in a single English adjective, and instead a paraphrase is needed. Second, it is necessary to compare images to reveal non-obvious attributes. Under comparison, the way in which people use non-obvious attributes to interpret images is systematic and highly stable. The images in the top row are ordered from left to right with respect to the extent to which the attribute "handle with care" applies. A given person may not agree that the description "handle with care" is applicable to the middle dog, but at the same time consider the ordering from left to right to be uncontroversial.

Ordering of images with respect to a non-obvious attribute represents an *interpretation dimension.* In the case of the dogs in Figure 1, this interpretation dimension stretches from "handle with care" to "companionable". It represents tacit knowledge of how to interpret dogs and dog images within the social community of people who take and share these social images. The interpretation dimension comes to light when images are compared, since comparison provides people with clues that trigger them to think of the images in terms of context and function. Additional examples and discussion of interpretation dimensions that people apply in understanding social images can be found in [4]. Here, we define an interpretation dimension as: a cluster of related non-obvious attributes that reflect a certain perspective on images used for interpretation. The caption of Figure 1 provides example interpretation dimensions for the 'livingroom' and 'tacos' images in the middle and bottom rows.

The goal of this paper is to present a procedure to discover non-obvious attributes that people use when interpreting social images. Ultimately, these non-obvious attributes and the associated interpretation dimensions will support the development of multimedia technology, i.e., classifiers and retrieval systems, that use representations reflecting, at a fined-grained level, the aspects of images that are important for human interpretation.

This paper is structured as follows. In the next section, we discuss non-obvious attributes and interpretation dimensions in more detail, comparing and contrasting them with related work. We then present our procedure for discovering non-obvious attributes using a particular example from the domain of fashion. Social fashion images are increasingly common on the Internet, and have attracted attention recently in the multimedia community, e.g., [5]. It is evident that much is to be gained if fashion can be described in terms of attributes that go beyond the obvious (e.g., basic concepts dress, shirt, tie) to capture the aspects of fashion images that are truly important for users. Next, we present the findings of our experiments in terms of a set of non-obvious attributes. Finally, we validate our findings by demonstrating that the non-obvious attributes that we discovered go beyond what can be derived from another source of information about user image interpretations, namely, user tags. We conclude with discussion.

## 2. BACKGROUND AND RELATED WORK

Recently a growing amount of research in the area of content-based image analysis has been directed at developing mid-level representations of images, i.e., attributes. In this section, we present two approaches related to ours, and discuss the unique contribution of our notion of non-obvious visual attributes.

### 2.1 Discovering image attributes with tags

The Visual Sentiment Ontology [1] consistes of mid-level features for describing images. The ontology was created by mining adjective-noun pairs from the tags of social images assumed to have a strong affective aspect, and creating visual detectors for each pair. The resulting ontology is tailored to represent the aspects of images that correspond to users expressions of sentiment. Note that non-obvious attributes we propose here are not subsumed by attributes related to affect. For example, the leftmost dog in Figure 1 could also be associated with the emotion "fear". However, fear is not a general characterization of the image. This point can be understood by noting that it is improbable that either the person who took the picture or the person who uploaded the picture (perhaps the dogs' owner) is afraid of the dog. "Handle with care", however, succeeds in distinguishing the leftmost dog from the other dogs in general manner. The attributes in [1] are learned on the basis of tags. Our procedure goes beyond this approach in that it is designed to discover tacit knowledge. Such knowledge is used to interpret images, but is not explicitly expressed, and often unconscious.

### 2.2 Eliciting image attributes from the crowd

Our approach has several important differences with previous uses of the crowd to elicit image attributes. We discuss these with reference to [6], a paper that interactively builds a vocabulary of nameable attributes. Our work differs in three main respects. First, [6] makes the assumption that people can explicitly name the aspects of images most important for interpretation. Although many attributes are without doubt nameable, nameability is not strictly necessary for people to interpret images. For example, someone who never has heard of "minimalism", cannot name it, but can still order the "livingroom" images in the middle row of Figure 1. Second, like [1], the work in [6] places a major focus on detectability. In contrast, we keep the focus firmly on how people interpret images, and not automatic analysis. Third, [6] presents two images to the crowdworker with no context in which they should be interpreted. Crowdworkers are asked to report which property changes from left to right. The paper defines properties for the crowdworkers, i.e., "properties include characteristics such as color or layout or general feel, but should not be names of objects, scenes, or animals". In contrast, we do not suggest attributes, but rather ask crowdworkers describe their process of comparing images within a process. The next section provides more details on how we elicit attributes within a context.

## 3. DISCOVERY VIA CROWDSOURCING

Our procedure for discovering non-obivous attributes and interpretation dimensions take the form of a crowdtask that we design and publish to Amazon Mechanical Turk (AMT). The task goes above and beyond existing crowdsourcing approaches with respect to its combination of two aspects. These two aspects address the challenges that face the discovery of non-obivous concepts mentioned in the introduction. First, the task asks crowdworkers to answer questions about images within a specific, typical use scenario that specifies function (i.e. the purpose for which the content of the images or the images themselves are used). Contextualizing the images in this way prompts people to move beyond naming basic concepts that they can perceive in the image, to thinking about the aspects of those concepts that are particularly important within a typical use scenario. Second, it makes use of *triadic elicitation* [2][9]. This method helps capture tacit knowledge by presenting three elements and asking people to elaborate in which way two elements are similar but different from the third. The triadic elicitation method has been successfully used for knowledge elicitation in crowdsourcing tasks [11], but, to our knowledge, not in the area of multimedia.

In this section, we describe our procedure via the discussion of a concrete Human Intelligence Task (HIT) for AMT. As previously mentioned our domain is fashion. We chose this domain since our broader aim is to develop a fashion trend analysis application. This application will crawl images from Twitter and attempt to predict fashion trends by analyzing similarities. Initial designs for the application focused on basic concepts (e.g., dresses, ties, jackets) and obvious attributes (e.g., color). However, it was soon realized that these concepts and attributes were inappropriate because they failed to capture the essence of what makes images of trendy clothing similar or different in the eyes of users.

**Comparing real-life fashion pictures**

Imagine that you have just opened a little fashion store and that you are deciding on the types of clothing that you would like to sell there. Your goal is to attract a diverse group of shoppers and you would like to offer them a stylish collection to choose from. However, you also want your store to be unique: to offer something different from the latest trends available at the big, mainstream stores. You are already well acquainted with "high fashion": trends from fashion shows. Now, you want to look at social media as they inform you about which outfits and accessories may appeal to your customers. For that purpose, you decide to study a set of fashion images from Flickr, three of which you will see below.

The depicted fashion items are different in some respects, and similar in others. Since you are interested in understanding the details of the upcoming trends, paying careful attention to similarities and differences is very important to you. We ask you to look at the fashion items in the images below and to give us your opinion about the similarities and differences that you see.

Below you can see three examples of images that we consider to belong to the category "*blouses*"

Blouse A     Blouse B     Blouse C

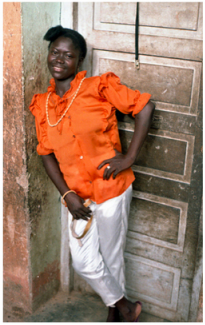 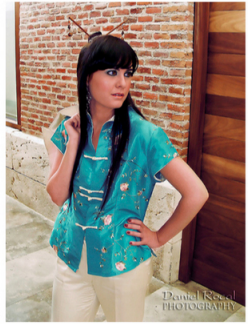 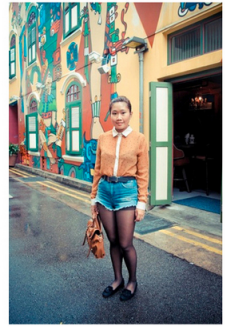

**Your task**

Please inspect the three images of blouses closely. Which statement is most applicable to the *blouses* depicted in this set of images?
○ The three blouses are all similar
○ The three blouses are all different
○ One of the blouses is different from the other two

Please write 2-3 sentences that tells us why you chose the answer. They should explain which aspects of the clothing or fashion accessories contribute to your opinion that the blouses are the same, and which aspects contribute to your opinion that they are different. Your answers should be focused on the *blouses*. That is, you should only pay attention to the Blouse in the image, and ignore everything else about the image. Particular characteristics of the people (for example: skin shade, hair style, body type) or of the background (for example: outdoors or indoors) are therefore **not** relevant. Please make clear references to the fashion items in your answers (e.g. Blouse A, B, C).

**Figure 2: Crowdsourcing task used to elicit interpretation dimensions, from which non-obvious attributes are extracted**

Previously we have presented a very brief overview of this task [blinded for review], and here we discuss it in the necessary depth needed to replicate the procedure. This overview is valuable because the set up of qualitative crowdsourcing experiments that have the purpose of revealing cognitive structures that influence users' interpretation of multimedia content is relatively uncovered territory. While an increasing number of studies assess the design, reliability and validity of surveys on AMT (e.g. [8][2][5]), little methodological guidance is available for qualitative research, such as the present case study.

### 3.1 Crowdsourcing Task Design

The design of the crowdsourcing task is displayed in Figure 2. The HIT starts with presenting a scenario in which workers are invited to imagine that they had just opened a little fashion store and were looking for types of clothing to be sold in the store. In the description, we provide information on the kind of shop, the clothing style, the goal of the shop owner (e.g., "*your goal is to attract a diverse group of shoppers and you would like to offer them a stylish collection to choose from*"), and the shop owner's prior knowledge (e.g., "*you are already well acquainted with 'high fashion': trends from fashion shows*"). The scenario states that the shop was using Flickr images in order to better understand "*which outfits and accessories may appeal to your customers*".

Crowdworkers are then presented with three fashion images. They are asked to indicate whether all images are similar, all images are different, or if one image differs from the other two. If the latter case holds, they are also asked to indicate which image was different from the other two. After answering this question, they are asked to explain their answer by elaborating on which aspects of the clothing or fashion accessories contributed to their opinion. We refer to this explanation as the "comparative description". It is from this description that we later extract the attributes. The HIT explicitly asks the crowdworkers to disregard contextual elements (e.g., background, kind of model) and focus on the fashion items.

The last part of the HIT contains information regarding the non-profit research project where data would be used, a link to the project website and a short description on the data handling procedure. Workers are explicitly informed that by accepting the HIT they are agreeing on the described data handling procedure.

### 3.2 Methodological considerations

The design of this crowdsourcing task was challenging for two reasons. The first is from the workers' side. The HIT asks a very open ended question. Crowdworkers are often used to carrying out HITs for which the requester is looking for one, specific, correct answer. For this reason, it is difficult for them to be confident that they are providing the types of descriptions that we, the requesters, will find satisfactory. Often they are concerned that we will reject their work.

In order to give workers confidence that they were providing the types of answers we were looking for, we introduced a "Reject List". This list constitutes a design innovation that we have not seen in another crowdsourcing HIT. The "Reject List" is a running list of the answers that workers submited that we rejected. This list was initially empty and was filled in as the HIT ran. The list helped the crowdworkers to understand what we were looking for. Note that the "Reject List" is not a substitute for detailed guidelines for the worker. The HIT also included a section containing explicit information on how to formulate the comparative descritions. We also included a small feedback section in which workers could point out issues with the HIT or propose suggestions on how to make it more enjoyable.

Further, in order to ensure that the workers understood the task, we applied the iterative design process proposed in [4] for eliciting interpretation dimensions from the crowd. Our goal was to support crowdworker's confidence in their own work by eliminating any "rough edges", i.e., respects in which the HIT fell short of being completely user-friendly.

The second challenge is from the side of the requester. The ultimate goal of this study was to get a rich set of answers which would allow us to formulate interpretation dimensions that reflect the fashion item properties. With this goal in mind, researchers have to set up a clear and, as much as possible, objective rejection criteria since rejections can introduce a researcher bias in the results: disallowing data to be included in the results can lead to the exclusion of certain perceptions of fashion items.

Following this reasoning, HITs were rejected only if they did not provide any information about similarities or differences between fashion items. This includes cases in which:

- the box for the comparative description is left empty,
- the comparative description only rephrases the selected option (Example: *"The three are looking different"*),
- the comparative description reflects a personal opinion about the fashion items, but does not point out similarities or differences between them (Example: *"Cool! everything is fabulous. I like all of them very much."*),
- the comparative description does not relate to the fashion items, but to other properties of the image or the persons depicted in the image (Example: *"Hair style, skin shade, body type"*).

During the HIT, we did not introduce any additional criteria for rejection. The "Reject List" accumulated only six answers: we added an answer each time we rejected a HIT that failed with respect to the guidelines. One of the answers was actually one that we accepted, but wanted to point out that we considered it borderline. We note that we have no tangible evidence that the "Reject List" improved the quality of the results we gathered, but none of the crowdworkers raised any objections in the comment box either, and, in general we consider it to be a success.

### 3.3 Crowdsourcing task execution

Because we wanted to collect the views of a large number of people each HIT was carried out by 33 crowdworkers. Each HIT contained one triad of images. In total, 37 triads were constructed using, in total, 111 Creative Commons (CC) fashion images that were retrieved manually from Flickr. The data we collected were thus collected from a total of 1221 HIT assignments. The images were selected to belong to one of 12 fashion categories (i.e., blouse, business suit, coat, dress, hipster, hoodie, nerd, poloshirt, pullover, running outfit, T-shirt, and trousers). These categories were chosen to represent the spectrum from common to contemporary, including fashion items as well as styles. For the purpose of this HIT, the search-based selection of the image triads was carried out by a single subject recruited by the authors on the basis of interest in and knowledge of fashion. We note that in the future, the selection of the images could also be carried out using a crowdsourcing task.

The AMT HIT took three days to complete. In total, 92 assignments (7.0%) were rejected. Rejected assignments were re-assigned. In total, 1313 HITs were submitted: 37 triads x 33 assignments per HIT + 92 rejected HITs. The average completion time for the approved HITs was 148.4 seconds (s.d. 123.0 seconds).

### 4. RESULTS

The comparative descriptions provided by the workers contained two/three sentences of natural language text. In order to go from these descriptions to attributes and from attributes to interpretation dimensions, we performed thematic analysis [9] on the data. The thematic analysis was carried out as follows: for each clothing category, the fashion-related expressions (words and phrases) which contained crowdworkers' textual answers were annotated (technically, the process is referred to as *coding*). The resulting inventory represents non-obvious image attributes related to fashion. Attributes were then thematically clustered into interpretation dimensions. The attributes and dimensions were documented in a coding scheme.

The analysis continues until the saturation point has been reached. This point (called *theoretical saturation*) is reached when adding new data to the analysis does not lead to further refinements of the coding scheme. This data analysis results in a phenomenological description of perceptions of fashion images in the form of interpretation dimensions, rather than in a set of deterministic class labels, such as is used for training conventional classifiers.

The answers returned by the crowdworkers for 1221 HITs were divided into 10 equal groups of 122 HITs. HITs were randomly assigned to one of the groups. We subsequently coded the HIT groups until theoretical saturation was achieved. This was the case after we analysed three groups, which corresponds to 30% of the 1221 HITs. As a second step, the resulting coding scheme was discussed between all authors and subsequently refined. Finally, the codes were manually clustered into interpretation dimensions. In the remainder of this section, we present the attributes that were discovered, and also the interpretation dimensions.

### 4.1 Fashion attributes from the crowd

In general, the crowdsourcing task yielded meaningful answers and well-elaborated argumentations. The quality of the input can be explained by the fact that the design of the HIT made it clear that we would be checking the input carefully. This aspect apparently motivated people who were interested in and knowledgable about clothing and fashion, but also discouraged people who were not. Many workers were engaged with the task, going beyond what was required in the description. Some of them provided long and very well argumented answers. We collected a rich variety of attributes, ranging from common (e.g., colour, length, material) to more unconventional (e.g. neck, button style). Table 1 illustrates exemplary answers and coded attributes for the clothing category "Blouse".

**Table 1: Illustration of how attributes (words and phrases) were extracted from the comparative descriptions contributed by the crowdworkers.**

| Example answers | Attributes |
|---|---|
| *"Blouse A is long sleeve, Blouse B is sleeveless, Blouse C is also sleeveless but has closed neck."* | Longsleeve, Sleeveless |
| *"All three blouses are similar, blouses A,B, and C have a bow tie at the collar."* | Collar, Bow neck |
| *"Blouse A is also a vibrant color while Blouse B and C are paler."* | Color vibrance |
| *"All three are women's blouses with fabric details by the collar (bow tie type fashion or ruffles)."* | Ruffles, Bow tie |
| *"The blouse shown in picture B is a casual one which can be used for everyday purpose."* | Casual |
| *"All three blouses seem to be made of a light, airy material."* | Light/airy |

Some of the identified attributes referred to quantifiable qualities such as color vibrance and length; whereas some others referred to categorical qualities, such as pattern, style, or situation in which the clothing is worn. Although most of the attributes can be generalized across fashion categories, a few people mentioned attributes that are specific for a fashion category, such as sleeve style for blouses; or specific for the picture, such as how to be worn (e.g., "*Blouse B and C are tucked into the women's shorts or skirts*").

Note that we are not particularly interested in the question of which of these attributes should be considered an "obvious" attribute, and which should be considered a "non-obvious" attribute. Instead, the key point is that the attributes discovered in this way would not have been revealed by a process that did not contextualize the images, or did not encourage people to compare them. We return to present evidence of this point in Section 5.

### 4.2 Attributes to interpretation dimensions

We carried out an interpretive process involving manual clustering of related attributes in order to identify interpretation dimensions. Within our specific use scenario, the fashion trend analysis application mentioned above, the dimensions will be used to sort images in the application. Note that in order to fulfill this goal we do not need the dimensions to provide perfect coverage of a semantic space. Rather they should provide insight into the data that goes above and beyond the obvious attributes commonly used for this purpose, e.g., color.

In Table 2, we provide an example of how the interpretative process goes from attributes to interpretation dimensions for the fashion category 'blouse'. The table demonstrates that our crowdsourcing-based procedure is capable of revealing interpretation dimensions corresponding to aspects important to users in this particular context. These aspects would be difficult to predict without consulting the crowd.

**Table 2 Interpretative process going from attributes to dimensions**

| Exemplary attributes | Dimension for the application |
| --- | --- |
| Long sleeves, Sleeveless, Cap sleeves, Short sleeves, Elbow length, Quarter sleeve, Close to sleeveless | Sleeve style |
| Collar/non-collar, Bow neck, Closed/open, Flexible collar/Stiff collar | Neck type |
| Vibrance/paleness, Dainty colors Neutral, Bright, Off-white | Color vibrance |
| Ruffles, Bow tie, Frill type, Accent lines parallel to the buttons. | Stitching |
| Casual, Asian, Churidhar-like, Formal, Dressy | Casualness |
| Light/airy, Satiny material, Wool material, Cotton material | Material |

### 5. ANALYSIS OF ADDED VALUE

In this section, we carry out a comparison of the attributes that we have collected with our crowdsourcing procedure to tags that users' assign to social images. The purpose of the analysis is to investigate whether our procedure is indeed discovering "non-obvious" aspects of images that go above an beyond what can already be gathered from explicit descriptions. In our experiment, we choose tags to be representative of explicit descriptions, since they are often used to mine semantics from social multimeida collections. Because the analysis is labor intensive, we focus on three of our 12 fashion categories: blouses, business suits, and trousers.

**Table 3 Statistics of tags associated with the images**

| Fashion category | Average no. of tags per image | % of fashion-related tags | % of tags related to category |
| --- | --- | --- | --- |
| Blouses | 28.9 | 45.2 | 22.8 |
| Business suits | 8.8 | 35.1 | 26.7 |
| Trousers | 17.8 | 31.1 | 18.2 |

Table 3 presents a summary of information about the tags. Many of the tags associated with the images were related to fashion. A smaller, but still significant percentage were associated with the fashion category of the image (i.e. specific to blouses, business suits, or trousers). Tags that were fashion related, but not category related, contained general fashion descriptors (e.g., fashion, outfit) and also were related to other fashion items depicted in the images (e.g., buckle, Rayban). A significant share of the non-fashion tags included tags that describe the location the picture was taken (e.g., California, house).

The core of our analysis is a qualitative comparison of the tags with the attributes that we discovered via our crowdsourcing-base procedure. Our first insight was that the crowdsourcing-discovered attributes were a super-set of those mentioned in the fashion tags. Next, our comparison revealed important differences with respect to the level of detail with which the users described the fashion items depicted in the images. We provide an example of the business suit in Figure 3.

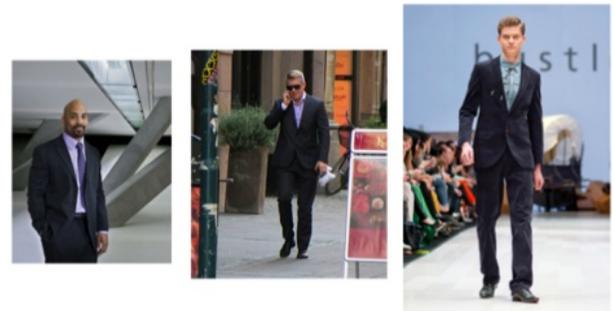

**Figure 3: Example of triad of "business suit" images**

The fashion-related tags of these three business suit images include *trousers, slacks, man in suit, suit, suited,* and *tie*. The tags denote the object in the image in different wordings (i.e., trousers and slacks). In contrast, the crowdsoucring-based attributes point to particular properties of those objects. The attributes include the fit, the design of the tie, the quality of the fabric, and the target group the suit is designed for. An example of a relevant comparative description written by a crowdworker is: *"Business suits [middle] and [right] are representative of the contemporary trend of tighter, form-fitting clothing; the jackets are buttoned up and possibly tailored to the size of the individual wearers. Business suit A is more traditional, looser-fitting and commonly worn unbuttoned as depicted."*.

In the case of the trousers, we uncovered a more subtle distinction. The tags do contain properties about the trousers, but these properties are less fine grained than the crowdsourcing-based attributes. For instance, some tags refer to the trousers' cut (e.g., *tight skinny jeans, tight*). In contrast, the attributes we found contain four different ways of expressing tightness, four general cut descriptors, and four different styles for the leg bottom. This granularity difference can be illustrated by the following comparative description: *"Trouser a and b has a more edgy look and feel. They are skinny and hug the leg from top to bottom which gives a sexy appearance as well. Trouser c has small flare at the bottom, but are basically straight leg and can be worn as business casual if need be."*

A similar pattern can be found for the blouses. A tag related to the sleeve style is 'longsleeve', whereas the descriptions gathered by the crowdsourcing task output distinguish between *cap sleeves*, *long sleeves*, *short sleeves*, *sleeveless*, *quarter sleeve*, and *close to sleeveless*. The following example is representative *"Blouses B and C are similar in that they are both light colored, with cap sleeves and a bow at the neck. Blouse A is quite different from the other two because it is brightly colored and has elbow-length sleeves and ruffles."*

This qualitative analysis is clearly limited in scope. However, it provides evidence that the use of the crowdsourcing-base procedure that we propose here supports the discovery of attributes and interpretation dimensions that are important for image interpretation, but that are not always explicitly expressed.

## 6. DISCUSSION AND CONCLUSION

We have presented the notion of non-obvious attributes and a procedure by which they can be discovered using the crowd. We have illustrated the procedure with an example drawn from fashion, a domain in which we have observed a need to move beyond readily recognizable attributes of images in order to arrive at aspects that are important for users. We have introduced interpretation dimensions, as clusters of non-obvious attributes. In this section, we discuss the ways in which the perspective and the procedure presented here opens up new challenges for content-based image analysis.

First, we point out the larger aim of our work. Ultimately, the purpose of the attributes discovered by our procedure is to move forward the state-of-the-art of multimedia technology (classifiers and retrieval systems) that uses representations of images. Non-obvious attributes are able to represent, at a fined-grained level, the aspects of images that are important for human interpretation. Other work on mid-level representations of images such as [1] and [6] devotes much attention to whether concepts or attributes are detectable in images using automatic methods. Naturally, this consideration is an important one. However, we see the future as lying not with fulling automatic imaging indexing systems, but rather with hybrid human conventional computation (H2C2) approaches. Such approaches combine automatic image analysis with input from the crowd. We anticipate that automatic image analysis will continue to grow more powerful, especially in light of the increasing amounts of training data available on the Internet. However, if a system is able to access crowdworkers continuously during the indexing process, there is no need to focus exclusively on machine detectability. A H2C2 system is able to detect any attribute that a human is able to detect.

Second, we explain the usefulness of conceptualizing image descriptions in terms of relative positions along interpretation dimensions rather than discreet categories. Interpretation dimensions have two desirable and helpful properties. The first property is that they abstract away from the exact wording that is chosen by the crowdworkers. Multiple crowdtasks would be expected to produce the same dimensions although the specific word choice of the crowdworkers may be different. The second property is that each interpretation dimension represents a possible spectrum along which images can be ordered. As pointed out in the introduction, we do not necessarily expect people to agree on whether a given attribute appropriately describes a given image. However, we do expect them to largely agree about the ordering of images along these dimensions. This agreement reflects tacit knowledge, which is used to interpret images, but not overtly expressed unless it is elicited. We point out that today's learning to rank approaches are able to exploit orderings and do not need traditional deterministic labels.

The major challenge that we see facing the exploitation of non-obvious attributes for image analysis is discovering exactly which interpretation dimenions are the ones that are most important and productive to pursue. We do not claim to have discovered every possible interpretation dimension with our approach. Nor do we feel that exhaustive inventory of non-obvious attributes is necessary. Instead, researchers should concentrate resources on those that are more helpful. Additional insight will be gained from repeating this experiment, and similar experiments to understand the stability of the dimensions that emerge.

## 7. ACKNOWLEDGMENTS

This work was partially funded by the European Commission's 7th Framework Programme under grant agreement no. 287704 (CUbRIK) and also by the Dutch National COMMIT program. We thank W. Ansah for her help with the exerpiment.